\documentclass[10pt,a4paper,onecolumn]{article}
\usepackage[a4paper, total={7.5in, 10.3in}]{geometry}
\usepackage{setspace}
\usepackage[hyphens]{url}
\usepackage{lineno,hyperref,multirow,cite}
\usepackage{authblk}
\usepackage{epsfig,amssymb,amsmath}
\usepackage[utf8]{inputenc}
\usepackage{amsmath,graphicx}
\usepackage{amssymb,amsmath,epsfig,url,mathrsfs} \usepackage{algorithm,algorithmic}
\usepackage[usenames, dvipsnames]{color}
\usepackage[framemethod=TikZ]{mdframed}
\usepackage{xcolor}
\usepackage{colortbl}
\usepackage[colorinlistoftodos]{todonotes}
\usepackage{multirow}
\usepackage{multicol}
\usepackage{enumitem}
\usepackage[margin=1cm]{caption}
\usepackage{longtable}
\usepackage{hhline}
\usepackage{numprint}
\npthousandsep{,}

\usepackage{graphicx}
\usepackage{subfig}
\usepackage{array, makecell}
\usepackage{verbatim}
\usepackage[utf8]{inputenc}
\usepackage[T1]{fontenc}
\usepackage[ngerman,english]{babel}
\usepackage{tabularx}  
\usepackage{ragged2e}  
\newcolumntype{Y}{>{\RaggedRight\arraybackslash}X} 
\usepackage{booktabs}  
\usepackage{amsthm}
 
\usepackage{tcolorbox}
\usepackage{graphicx}
 
\definecolor{chestnut}{rgb}{0.8, 0.36, 0.36}
\definecolor{frenchblue}{rgb}{0.0, 0.45, 0.73}
\definecolor{babyblue}{rgb}{0.54, 0.81, 0.94}
\definecolor{beaublue}{rgb}{0.74, 0.83, 0.9}
\definecolor{bubblegum}{rgb}{0.99, 0.76, 0.8}
\definecolor{gainsboro}{rgb}{0.86, 0.86, 0.86}
\definecolor{amethyst}{rgb}{0.6, 0.4, 0.8}
\definecolor{mygray}{gray}{0.85}
\definecolor{aq}{rgb}{0.5, 1.0, 0.83}
\definecolor{aquamarine}{rgb}{0.5, 1.0, 0.83}
\definecolor{beaublue}{rgb}{0.74, 0.83, 0.9}
\definecolor{cadetblue}{rgb}{0.37, 0.62, 0.63}
\definecolor{lavender}{rgb}{0.9, 0.9, 0.98}
\definecolor{nonphotoblue}{rgb}{0.64, 0.87, 0.93}
\definecolor{paleaqua}{rgb}{0.74, 0.83, 0.9}
\definecolor{peachpuff}{rgb}{1.0, 0.85, 0.73}
\definecolor{uscgold}{rgb}{1.0, 0.8, 0.0}
\definecolor{pastelorange}{rgb}{1.0, 0.7, 0.28}
\definecolor{mountainmeadow}{rgb}{0.19, 0.73, 0.56}
\definecolor{mossgreen}{rgb}{0.68, 0.87, 0.68}
\definecolor{lemonchiffon}{rgb}{1.0, 0.98, 0.8}
\definecolor{lightblue}{rgb}{0.68, 0.85, 0.9}
\definecolor{azuremist}{rgb}{0.94, 1.0, 1.0}
\definecolor{caribbeangreen}{rgb}{0.0, 0.8, 0.6}
\definecolor{celadon}{rgb}{0.67, 0.88, 0.69}
\definecolor{camouflagegreen}{rgb}{0.47, 0.53, 0.42}
\definecolor{coolgrey}{rgb}{0.55, 0.57, 0.67}
\definecolor{darkolivegreen}{rgb}{0.33, 0.42, 0.18}
\definecolor{darkspringgreen}{rgb}{0.09, 0.45, 0.27}
\definecolor{beige}{rgb}{0.96, 0.96, 0.86}
\definecolor{lightapricot}{rgb}{0.99, 0.84, 0.69}
\definecolor{navajowhite}{rgb}{1.0, 0.87, 0.68}
\definecolor{pastelorange}{rgb}{1.0, 0.7, 0.28}

\makeatletter
\newcommand{\mathleft}{\@fleqntrue\@mathmargin0pt}
\newcommand{\mathcenter}{\@fleqnfalse}
\makeatother
 
\theoremstyle{definition}

\usepackage{todonotes}

\hypersetup{
   unicode=false,          
   pdftoolbar=true,        
   pdfmenubar=true,        
   pdffitwindow=false,     
   pdfstartview={FitH},    
   pdftitle={VoicePrivacy 2024 Challenge Evaluation Plan},    
   pdfauthor={VoicePrivacy},     
   pdfsubject={VoicePrivacy},   
   pdfcreator={},   
   pdfproducer={}, 
   pdfkeywords={VoicePrivacy, anonymisation}, 
   pdfnewwindow=true,      
   colorlinks=true,       
   linkcolor=blue,          
   citecolor=blue,        
   filecolor=magenta,      
   urlcolor=blue           
}

\usepackage{color, colortbl}
\definecolor{LightCyan}{rgb}{0.88,1,1}

\mdfdefinestyle{MyFrame}{%
    linecolor=black,
    outerlinewidth=.1pt,
    roundcorner=1pt,
    innertopmargin=\baselineskip,
    innerbottommargin=\baselineskip,
    innerrightmargin=5pt,
    innerleftmargin=-1pt,
    backgroundcolor=gray!10!white}

\title{\textbf{{The  First VoicePrivacy  Attacker  Challenge Evaluation Plan\\[1em]\large{}Version 2.1}}}

\author[1]{Natalia Tomashenko}
\author[2]{Xiaoxiao Miao}
\author[1]{Emmanuel Vincent}
\author[3]{Junichi Yamagishi}

 \affil[1]{Inria, France}
 \affil[2]{Singapore Institute of Technology, Singapore}
 \affil[3]{National Institute of Informatics, Tokyo, Japan}

\date{\url{https://www.voiceprivacychallenge.org/attacker/} \\
 \href{mailto:attacker.challenge@inria.fr}{attacker.challenge@inria.fr}}
\begin{document}

\maketitle
\normalsize

 \abstract
\normalsize
The First VoicePrivacy Attacker Challenge is a new kind of challenge organized as part of the VoicePrivacy initiative \cite{tomashenko2020introducing} and supported by \textbf{ICASSP 2025} as the \textbf{SP Grand Challenge}.\footnote{\url{https://2025.ieeeicassp.org/sp-grand-challenges/\#gc7}}
It focuses on developing \textbf{attacker systems against voice anonymization}, which will be evaluated against a set of anonymization systems submitted to the VoicePrivacy 2024 Challenge\footnote{The VoicePrivacy Challenge focuses on strengthening voice anonymization systems from the user's perspective, often assuming fixed attack scenarios, which may not fully reflect practical use cases, as real-world attacks can exploit any available clues and resources. In contrast, the Attacker Challenge aims to develop more robust and practical attacker systems capable of challenging various advanced anonymization systems from the attacker's point of view.}. 
Training, development, and evaluation datasets are provided along with a baseline attacker system. Participants shall develop their attacker systems in the form of automatic speaker verification systems and submit their scores on the development and evaluation data to the organizers. To do so, they can use any additional training data and models, provided that they are openly available and declared before the specified deadline. The metric for evaluation is equal error rate (EER).
Results will be presented at the ICASSP 2025 special session to which 5 selected top-ranked participants will be invited to submit and present their challenge systems. 
 
\begin{tcolorbox}[width=0.91\textwidth, colback={navajowhite}, title={\textbf{Changes in version 2.1 w.r.t. 2.0}}, colbacktitle=pastelorange, coltitle=black, arc=0.3mm, fonttitle=\bfseries, boxrule=0.5pt]  
 
Updated  list of data and models to train  attacker systems (Table~\ref{tab:data-models-final-list}).
\end{tcolorbox}

\section{Context} 

\normalsize
  
Speech encapsulates a wealth of personal, private data, e.g., age and gender, health and emotional state, racial or ethnic origin, geographical background, social identity, and socio-economic status \cite{Nautsch-PreservingPrivacySpeech-CSL-2019}. 
Formed in 2020, the VoicePrivacy initiative \cite{tomashenko2020introducing} is promoting the development of privacy preservation solutions for speech technology via a series of
competitive benchmarking challenges, with common datasets, protocols and metrics. In this context, privacy preservation is classically formulated as a game between \emph{users} who process their utterances (referred to as \emph{trial} utterances) with a privacy preservation system prior to sharing with others, and \emph{attackers} who access these processed utterances or data derived from them and wish to infer information about the users. The level of privacy offered by a given solution is measured as the lowest error rate among all attackers. 

The first three VoicePrivacy Challenge editions
\cite{Tomashenko2021CSl,tomashenko2024voiceprivacy,Tomashenko2021CSlsupplementay,panariello2024voiceprivacy,tomashenko2022voiceprivacy,bonastre2021benchmarking} 
focused on the development of voice anonymization systems. In particular, the systems submitted to the VoicePrivacy 
2024 Challenge had to meet the following requirements: 
    (a) output a speech waveform; 
    (b) conceal the speaker identity at the \emph{utterance level}; 
    (c) not distort the linguistic and emotional content.
The processed utterances sound as if they were uttered by another speaker, which we refer to as a \emph{pseudo-speaker}. The pseudo-speaker is selected independently for every utterance, and it can be an artificial voice not corresponding to any real speaker. In practice, many voice anonymization systems select the pseudo-speaker or modify prosody in a random or semi-random way using a random number generator. A \emph{semi-informed attack model} \cite{srivastava2019evaluating} was assumed, whereby attackers have access to the voice anonymization system (but not to the random numbers drawn by that system for each utterance, if any), and they seek to re-identify the original speaker behind each anonymized trial utterance. Specifically, an ECAPA-TDNN \cite{desplanques2020ecapa} automatic speaker verification (ASV) system was provided by the organizers and trained by the participants on data anonymized using their anonymization system. While this attack model is undeniably the most realistic to date, the provided attacker system is not its strongest possible implementation as it does not exploit spoken content similarities, specific pseudo-speaker selection strategies \cite{champion2021invertibility}, or stronger ASV architectures \cite{zeng2022attention}, among others.

To ensure a fair and reliable privacy assessment, it is essential to find the strongest possible attacker against every anonymization system. Hence, the current challenge edition takes the attacker's perspective and focuses on the development of attacker systems against voice anonymization systems.

\normalsize

\section{Task}
\label{sec:task}
Participants are required to develop one or more attacker systems against one or more voice anonymization systems selected among three VoicePrivacy 2024 Challenge baselines \cite{tomashenko2024voiceprivacy} and four
systems developed by the VoicePrivacy 2024 Challenge participants. 
For each speaker of interest, the attacker is assumed to have access to one or more utterances spoken by that speaker, which 
are referred to as \textit{enrollment} utterances.
The attacker system shall 
output an ASV score for every given pair of trial utterance and enrollment speaker, where higher (resp., lower) scores correspond to same-speaker (resp., different-speaker) pairs.

To develop and evaluate their attacker system against a given voice anonymization system, in line with the assumed semi-informed attack model, participants have access to:
\vspace{-\topsep}
\begin{enumerate}[label=(\alph*)]
\setlength{\parskip}{0pt} \setlength{\itemsep}{0pt plus 1pt}
\item anonymized trial utterances;
\item 
original and
anonymized 
enrollment utterances;
\item 
original and
anonymized training data (as well as other publicly available training resources that will be specified in Section~\ref{sec:data}) for the ASV system;
\item a written description of the voice anonymization system;
\item the software implementation of that voice anonymization system when available. 
\end{enumerate}
\vspace{-\topsep}

\section{Data}\label{sec:data}

For each voice anonymization system, 
participants are provided with training, development and evaluation data anonymized using that system.
A detailed description of these datasets 
is presented below and in Table~\ref{tab:data}.

\begin{table}[!th]
\centering
  \caption{Number of speakers and utterances in the attacker training, development, and evaluation sets.}\label{tab:data}
 \resizebox{0.73\textwidth}{!}{
  \centering
  \begin{tabular}{|c|l|l|r|r|r|r|}
\hline
 \multicolumn{3}{|c|}{\textbf{Subset}} &  \textbf{Female} & \textbf{Male} & \textbf{Total} & \textbf{\#Utterances}  \\ \hline \hline
 \multirow{1}{*}{{Training~}} & \multicolumn{2}{l|}{ LibriSpeech: train-clean-360} & \numprint{439} & \numprint{482} & \numprint{921} & \numprint{104014} \\ \hline\hline
\multirow{2}{*}{{~ Development }} & LibriSpeech & Enrollment & 15 & 14 & 29 & 343\\ \cline{3-7}
& dev-clean & Trial & 20 & 20 & 40 & \numprint{1978}\\ \cline{1-7}
\multirow{2}{*}{{Evaluation~}} & LibriSpeech & Enrollment & 16 & 13 & 29 & 438\\ \cline{3-7}
& test-clean & Trial & 20 & 20 & 40 & \numprint{1496}\\ \cline{1-7}
\end{tabular}}
\end{table}
\normalsize

\paragraph{Training resources.} The training set is
the \textit{train-clean-360} subset of the \textit{{LibriSpeech}} \cite{panayotov2015librispeech} corpus. Besides the provided anonymized training data, participants are allowed to use the original \textit{train-clean-360} data.
In addition, participants were allowed to propose other resources such as speech corpora and pretrained models before the deadline (13th October).
Based on the  suggestions received from the challenge participants, in this version of the evaluation plan, we publish the final list of training data and pretrained   models  allowed for training attacker systems. All the allowed resources are listed in Table~\ref{tab:data-models-final-list}.

For some models, 
the provided link is a webpage listing multiple versions of the model. In this case, unless otherwise stated, all model versions available on that page before 15th October 2024 can be used by participants in  training their attacker systems.
Participants are allowed to use any existing software in the development and training of their attacker systems. If the software uses pretrained models, these models should be explicitly listed in this table or  on the main page (readme) of the corresponding repository before 15th October 2024.

\begin{longtable}[htbp!]{|c|p{3.4cm}|p{12cm}|}
  \caption{Final list of models and data for training attacker systems.}\label{tab:data-models-final-list}\\
  \hline
$ \textbf{\#}$	&	\textbf{Model}	&		\textbf{Link}	\\ \hhline{===}   
1	&	WavLM  	&	\url{	https://github.com/microsoft/unilm/tree/master/wavlm} \\
    &  Base~and~Large \cite{Chen2021WavLM} & \url{https://huggingface.co/microsoft/wavlm-large} \\ \hline 
2	&	Whisper \cite{radford2023robust}	&	\url{https://github.com/openai/whisper} \\
& & \url{https://huggingface.co/openai/whisper-large} \\ \hline
3	&	HuBERT \cite{hubert} 	&	\url{https://github.com/facebookresearch/fairseq/blob/main/examples/hubert} \\
& & \url{https://huggingface.co/facebook/hubert-large-ls960-ft} \\ \hline
4	&	XLS-R \cite{babu2021xls}	&	\url{	https://github.com/facebookresearch/fairseq/blob/main/examples/wav2vec/xlsr	} \\
& & \url{https://huggingface.co/facebook/wav2vec2-large-xlsr-53} \\ \hline
5	&	wav2vec 2.0 \cite{baevski2020wav2vec}	&	\url{	https://github.com/facebookresearch/fairseq/tree/main/examples/wav2vec	} \\
	&		&	\url{	https://dl.fbaipublicfiles.com/voxpopuli/models/wav2vec2_large_west_germanic_v2.pt	} \\
 & & \url{https://huggingface.co/facebook/wav2vec2-large-960h-lv60} \\ \hline
6	&	ECAPA-TDNN	\cite{desplanques2020ecapa} &	\url{	https://huggingface.co/speechbrain/spkrec-ecapa-voxceleb	} \\ \hline
7	&	NaturalSpeech 3	\cite{ju2024naturalspeech} &	\url{	https://huggingface.co/amphion/naturalspeech3_facodec	} \\ \hline

8	&	Encodec \cite{encodec}	&	\url{	https://huggingface.co/facebook/encodec_24khz	} \\ \hline
9	&	Bark	&	\url{	https://huggingface.co/suno/bark	} \\
	&		&	\url{	https://huggingface.co/erogol/bark/tree/main	} \\ \hline 
10 & Resnet34 &	\url{https://wenet.org.cn/downloads?models=wespeaker&version=voxceleb_resnet34.zip} \\ \hline
11 & Resnet34\_lm & \url{https://wenet.org.cn/downloads?models=wespeaker&version=voxceleb_resnet34_LM.zip} \\ \hline
12 & VGGVox &	\url{https://github.com/a-nagrani/VGGVox} \\ \hline
13 & Kaldi  models &	\url{http://kaldi-asr.org/models.html} \\ \hline

14 & Conformer models  &	\url{https://huggingface.co/models?search=conformer} \\ \hline

15	&	NVIDIA TitaNet-Large (en-US) \cite{koluguri2022titanet}	 &	\url{	https://huggingface.co/nvidia/speakerverification_en_titanet_large	} \\ \hline

 \multicolumn{3}{l}{}  \\ \hline
$ \textbf{\#}$		&	\textbf{Dataset}	&	\textbf{Link}	\\ \hhline{===}

16	&	LibriSpeech \cite{panayotov2015librispeech}: train-clean-100, train-clean-360, train-other-500	&	\url{	https://www.openslr.org/12	} \\ \hline

17	&	RAVDESS \cite{livingstone2018ryerson}	&	\url{	https://datasets.activeloop.ai/docs/ml/datasets/ravdess-dataset/} \\
	&		&	\url{	https://zenodo.org/records/1188976} \\ \hline

18	&	SAVEE \cite{haq2009speaker}	&	\url{	http://kahlan.eps.surrey.ac.uk/savee/} \\
& & \url{https://www.kaggle.com/datasets/ejlok1/surrey-audiovisual-expressed-emotion-savee        }	 \\ \hline

19	&	EMO-DB	\cite{burkhardt2005database}&	\url{http://emodb.bilderbar.info/download/} \\ \hline

20	&	VoxCeleb-1,2 \cite{chung2018voxceleb2}	& \url{https://www.robots.ox.ac.uk/~vgg/data/voxceleb/index.html\#about} \\ \hline

21	&	CMU-MOSEI	\cite{zadeh2018multimodal} &	\url{	http://multicomp.cs.cmu.edu/resources/cmu-mosei-dataset/	} \\ \hline

22	&	MUSAN \cite{snyder2015musan}	&	\url{	https://www.openslr.org/17/	} \\ \hline

23	&	RIR 	\cite{ko2017study} &	\url{	https://www.openslr.org/28/	} \\ \hline

24 & CN-Celeb \cite{li2023cn} &	\url{https://cnceleb.org/} \\ \hline

25 & Common Voice	\cite{ardila2019common} & \url{https://commonvoice.mozilla.org/en/datasets} \\ \hline

26 & IEMOCAP \cite{busso2008iemocap} & 	 \url{https://sail.usc.edu/iemocap/iemocap_info.htm} \\ \hline

27 & Emo-DB \cite{burkhardt2005database} & 	 \url{http://emodb.bilderbar.info/index-1280.html} \\ \hline

28 & Toronto Emotional Speech Database & 	 \url{https://tspace.library.utoronto.ca/handle/1807/24487} \\ \hline

29 & LIRIS-ACCEDE \cite{baveye2015liris} & 	\url{https://liris-accede.ec-lyon.fr/database.php} \\ \hline

30 & AESDD \cite{vryzas2018speech} & 	 \url{http://m3c.web.auth.gr/research/aesdd-speech-emotion-recognition/    } \\ \hline

31 & ANAD & 	 \url{https://www.kaggle.com/suso172/arabic-natural-audio-dataset} \\ \hline

32 & BAVED \cite{ali_aouf_2020} & 	 \url{https://www.kaggle.com/a13x10/basic-arabic-vocal-emotions-dataset} \\ \hline

33 & VoxForge & 	 \url{http://www.voxforge.org/} \\ \hline

34 & TED-LIUM 3 \cite{hernandez2018ted} & 	 \url{https://www.openslr.org/51/} \\ \hline

35 & AISHELL-1 \cite{bu2017aishell} & 	 \url{https://www.openslr.org/33/} \\ \hline

36 & AISHELL-WakeUp-1 &\url{https://www.aishelltech.com/wakeup\_data} \\ \hline

37 & CMU-MOSEI \cite{zadeh2018multimodal} & 	 \url{http://multicomp.cs.cmu.edu/resources/cmu-mosei-dataset/} \\ \hline

38 & VoxBlink2 \cite{lin24voxblink2} & 	 \url{https://voxblink2.github.io/} \\ \hline

39 & Emotional  Voices Database \cite{adigwe2018emotional}	 & \url{https://github.com/numediart/EmoV-DB} \\ \hline

40 & DEMoS \cite{parada2020demos} & 	 \url{https://zenodo.org/record/2544829} \\ \hline

\multicolumn{3}{l}{}  \\ \hline

 $\textbf{\#}$		&	\textbf{Software with pretrained models}	&	\textbf{Link}	 \\ \hhline{===}
41	&	VITS \cite{kim2021conditional}	&	\url{	https://github.com/jaywalnut310/vits/	} \\ 
	&		&	Models: \url{	 https://drive.google.com/drive/folders/1ksarh-cJf3F5eKJjLVWY0X1j1qsQqiS2	} \\ \hline
42 & SASV fusion-based baseline from the SASV 2022 challenge \cite{wang2020asvspoof} &	\url{https://github.com/sasv-challenge/SASVC2022_Baseline} \\ \hline
43 & Baseline of the ASVspoof 5 challenge Track 2, SASV	\cite{delgado2024asvspoof} & \url{https://github.com/sasv-challenge/SASV2_Baseline/tree/asvspoof5} \\ \hline 

44 & ASV-Subtools \cite{tong2021asv} &	\url{https://github.com/Snowdar/asv-subtools} \\ 
   \\ \hline
45 &   WeSpeaker  \cite{wang2023wespeaker} &	\url{https://github.com/wenet-e2e/wespeaker} \\ 
   &              & Models: \url{https://github.com/wenet-e2e/wespeaker/blob/master/docs/pretrained.md} 
   \\ \hline
46 & 3D-Speaker \textit{ERes2Net}~\cite{chen2023enhanced}
\textit{ERes2NetV2}~\cite{chen2024eres2netv2}
\textit{CAM++}~\cite{wang2023cam++} & \url{https://github.com/modelscope/3D-Speaker} \\ \hline

47 & SpeechBrain	\cite{speechbrain} & \url{https://speechbrain.github.io/} \\
   &                & Models: \url{https://huggingface.co/speechbrain} \\ \hline
48 & ResNet \cite{hu2018squeeze} & 	\url{https://github.com/ranchlai/speaker-verification} \\ \hline
49 & VQMIVC \cite{wang21n_interspeech} & 	\url{https://github.com/Wendison/VQMIVC} \\ \hline
50 & DeepSpeech \cite{hannun2014deep} & 	\url{https://github.com/mozilla/DeepSpeech} \\
   &                & Models: \url{https://github.com/mozilla/DeepSpeech/releases}\\ \hline
51 & AutoVC \cite{autovc2019} & 	\url{https://github.com/auspicious3000/autovc} \\ \hline
\end{longtable}
\normalsize

\paragraph{Development and evaluation data.}

 The development and evaluation sets comprise \textit{LibriSpeech dev-clean} and \textit{LibriSpeech test-clean}. 
 Besides the provided anonymized enrollment data, the participants are allowed to use the original enrollment data.

\paragraph{Voice anonymization systems.}\label{sec:anon_systems}
The voice anonymization systems to be attacked include three baseline systems (\textbf{B3}, \textbf{B4}, and \textbf{B5}) \cite{tomashenko2024voiceprivacy} and four selected 
systems developed by the VoicePrivacy 2024 Challenge participants (\textbf{T8-5}, \textbf{T10-2}, \textbf{T12-5}, and \textbf{T25-1}) \cite{tomashenko2024presentation}.
The participants' systems were chosen according to their anonymization performance in the highest privacy category ($\text{EER}\geq40\%$), excluding cascaded anonymization systems based on automatic speech recognition (ASR) followed by text-to-speech (TTS).
Thus, a total of seven systems are to be attacked:

\begin{itemize}
    \item \textbf{B3}  --  based on phonetic transcription, pitch and energy modification, and artificial pseudo-speaker embedding generation \cite{meyer2023prosody,tomashenko2024voiceprivacy}.
    
 \item \textbf{B4}  --  based on neural audio codec language modeling \cite{panariello_speaker_2023,tomashenko2024voiceprivacy}.

 \item \textbf{B5} --  based on vector quantized bottleneck (VQ-BN) features extracted from an ASR model and on original pitch \cite{champion2023,tomashenko2024voiceprivacy}.

  \item \textbf{T8-5}  (team \textit{JHU-CLSP}, system \textit{“Admixture ($p=0.4$)”} \cite{xinyuan2024hltcoe}) --  random selection of one of two methods for each utterance (with probability $p$ for the second method): (1) a cascaded ASR-TTS system with \textit{Whisper} \cite{radford2023robust} for ASR and \textit{VITS} \cite{kim2021conditional} for TTS  and (2) a k-nearest neighbor (kNN) voice conversion (VC) system operating on \textit{WavLM} \cite{Chen2021WavLM} features.

\item \textbf{T10-2}  (team \textit{NPU-NTU}, system \textit{“C4”} \cite{yao2024npu}) --  neural audio codec, with a specific disentanglement strategy  
for linguistic content, speaker identity and emotional state.

\item \textbf{T12-5} (team \textit{NTU-NPU}, system \textit{“3”} \cite{Kuzmin2024ntu}) -- based on \textbf{B5}, with
additional pitch smoothing.

\item \textbf{T25-1} (team \textit{USTC-PolyU}, system  \textit{“large: ESD+LibriTTS”} \cite{Gu2024ustc}) -- disentanglement of content (VQ-BN as in \textbf{B5}) and style (global style token (GST) \cite{wang2018style}) features
and emotion transfer
  from target speaker utterances.
 
\end{itemize}
\vspace{-\topsep}

%
The code of \textbf{B3}, \textbf{B4}, and \textbf{B5} is available on GitHub\footnote{\url{https://github.com/Voice-Privacy-Challenge/Voice-Privacy-Challenge-2024}} and can be used to develop attacker systems by, e.g., generating different or additional training data to train those systems.

\section{Evaluation metric}
\label{sec:metric}

We use the equal error rate (EER) metric to evaluate the attacker's performance. This metric has been used in all VoicePrivacy Challenge editions. 
For every given pair of trial utterance and enrollment speaker, the attacker system outputs an ASV score from which a same-speaker vs.\ different-speaker decision is made by thresholding.
Denoting by $P_\text{fa}(\theta)$ and $P_\text{miss}(\theta)$ the false alarm and miss rates at threshold~$\theta$, the EER metric corresponds to the threshold $\theta_\text{EER}$ at which the two detection error rates are equal, i.e., $\text{EER}=P_\text{fa}(\theta_\text{EER})=P_\text{miss}(\theta_\text{EER})$.
The lower this metric, the stronger the attacker.
The number of same-speaker and different-speaker trials in the development and evaluation datasets is given in Table~\ref{tab:trials}. The attackers will be ranked separately for each voice anonymization system.

\begin{table}[htbp]
  \caption{Number of speaker verification trials.}\label{tab:trials}
  \centering
   \resizebox{0.7\textwidth}{!}{
  \begin{tabular}{|l|l|l|r|r|r|}
\hline
 \multicolumn{2}{|c|}{\textbf{Subset}} & \textbf{Trials} &  \textbf{Female} & \textbf{Male} & \textbf{Total}  \\ \hline \hline
\multirow{2}{*}{{Development~}} & LibriSpeech & Same-speaker & 704 & 644 & \numprint{1348} \\ \cline{3-6}
 & dev-clean & Different-speaker	& \numprint{14566} & \numprint{12796} &	\numprint{27362} \\ \cline{1-6}
\multirow{2}{*}{{Evaluation~}} & LibriSpeech & Same-speaker & 548 & 449	& \numprint{997} \\ \cline{3-6}
  & test-clean & Different-speaker & \numprint{11196} & \numprint{9457} &	\numprint{20653} \\ \cline{1-6}
  \end{tabular}}
\end{table}

\section{Baseline attacker system}\label{sec:baseline_attack}

\begin{figure}[h!]
\centering
\includegraphics[width=175mm]{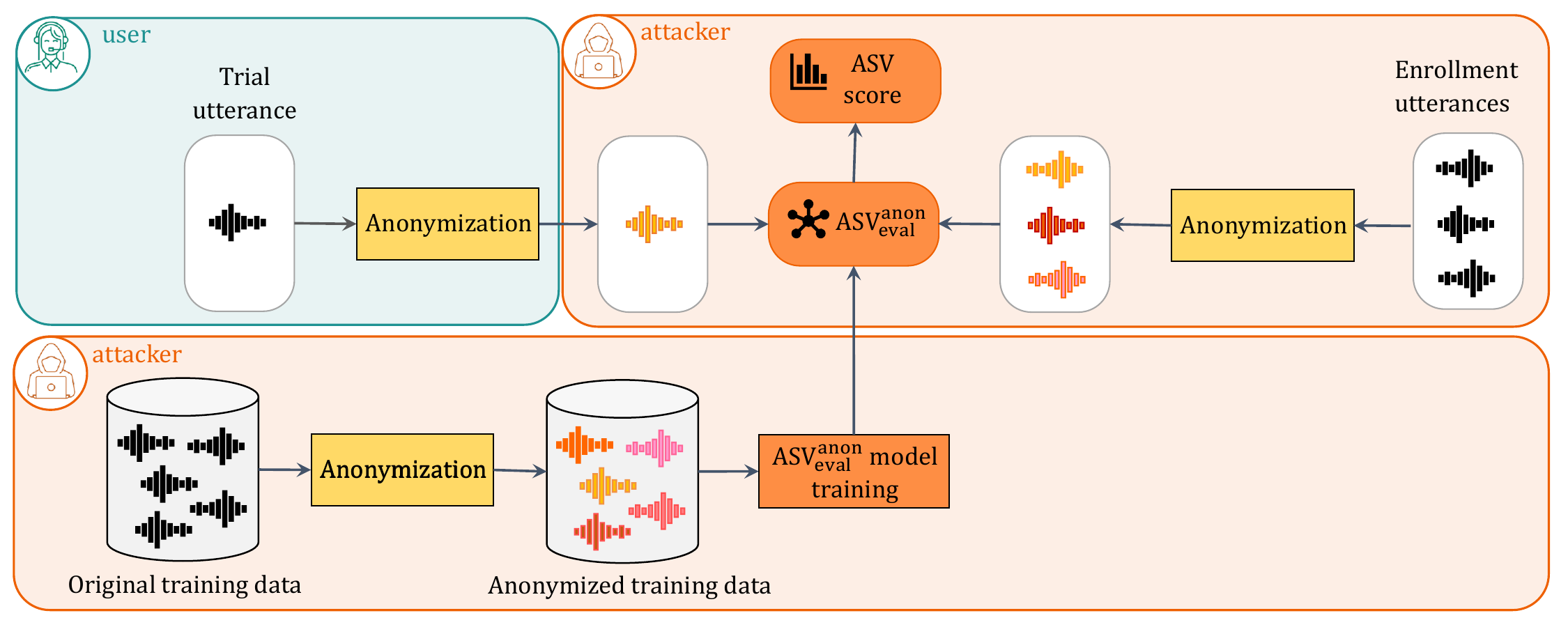}
\caption{Baseline attacker: training $ASV_\text{eval}^{\text{anon}}$ on anonymized training data and using it to compare anonymized trial and enrollment data.}
\label{fig:asv-eval}
\end{figure}

As a baseline, we consider the attacker system used in the VoicePrivacy 2024 Challenge \cite{tomashenko2024voiceprivacy} (see Figure~\ref{fig:asv-eval}).
The ASV system (denoted $ASV_\text{eval}^{\text{anon}}$) is an ECAPA-TDNN \cite{desplanques2020ecapa} with 512 channels in the convolution frame layers, implemented by adapting the \textit{SpeechBrain} \cite{speechbrain} \textit{VoxCeleb} recipe to \textit{LibriSpeech}, and it is trained on  anonymized training data.
For a given trial utterance and enrollment speaker, the attacker computes the average speaker embedding of all anonymized enrollment utterances from that speaker and compares it to the speaker embedding of the anonymized trial utterance.
Results for this baseline attacker system are reported in Table~\ref{tab:asv-results}.
The code to train the baseline attacker systems for given anonymized data is available in the GitHub  VoicePrivacy 2024 Challenge repository: \url{https://github.com/Voice-Privacy-Challenge/Voice-Privacy-Challenge-2024}.\footnote{See \textit{Step 2: Evaluation, \href{https://github.com/Voice-Privacy-Challenge/Voice-Privacy-Challenge-2024/blob/main/run_evaluation.py}{run\_evaluation.py}}}

\begin{table*}[h]
   \caption{EER achieved by the baseline attacker system $ASV_\text{eval}^\text{anon}$ on data processed by different anonymization systems vs.\ EER achieved 
  on the original (Orig.) unprocessed data by an ASV model  trained on original data.}\label{tab:asv-results}
  \centering
  \begin{tabular}{|l|l||c||r|r|r|r|r|r|r|}
\Xhline{0.7pt}
 \multirow{2}{*}{	\textbf{Dataset}} &  \multirow{2}{*}{\textbf{Gender}}  & \multicolumn{8}{c|}{\textbf{EER (\%)}}\\ \cline{3-10}
&  &{	\textbf{Orig.}}  &  \multicolumn{1}{|c|}{\textbf{B3}} &  \multicolumn{1}{|c|}{\textbf{B4}} &  \multicolumn{1}{|c|}{\textbf{B5}} &  \multicolumn{1}{|c|}{\textbf{T8-5}} & \multicolumn{1}{|c|}{\textbf{T10-2}} & \multicolumn{1}{|c|}{\textbf{T12-5}}  & \multicolumn{1}{|c|}{\textbf{T25-1}}\\ \hline \hline
\multirow{2}{*}{LibriSpeech-dev}    &   female  &	10.51 & 	 	28.43 & 	34.37 & 	35.82 & 	39.63 & 43.63 & 43.32 & 42.65\\
                                    &	male    &	0.93 & 	    	   22.04 & 	   31.06 & 	    32.92 & 	40.84 & 40.04 & 44.10 & 40.06\\ \hline
\multicolumn{2}{|c||}{\cellcolor{gray!7} Average dev} &		\cellcolor{gray!7} 5.72	  	& \cellcolor{gray!7} 25.24 & \cellcolor{gray!7} 32.71 & \cellcolor{gray!7} 34.37 & \cellcolor{gray!7}40.24 & \cellcolor{gray!7} 41.83 & \cellcolor{gray!7} 43.71 & \cellcolor{gray!7} 41.36 \\ \hline\hline
\multirow{2}{*}{LibriSpeech-test}	&	female 		&		8.76  & 	 	27.92 & 	29.37 & 	33.95 & 	42.50 & 41.97 & 43.61 & 42.34\\
	                                &	male 		&		0.42 & 	 	26.72 & 	31.16 & 	34.73 & 	40.05 & 38.75 & 41.88 &  41.92 \\  \hline
 \multicolumn{2}{|c||}{\cellcolor{gray!7} Average eval} &		\cellcolor{gray!7}4.59	 & \cellcolor{gray!7} 27.32 & \cellcolor{gray!7} 30.26 & \cellcolor{gray!7} 34.34 & \cellcolor{gray!7} 41.28 & \cellcolor{gray!7} 40.36 & \cellcolor{gray!7}42.75 & \cellcolor{gray!7} 42.13 \\
\Xhline{0.7pt}
  \end{tabular}  
\end{table*}

\section{Evaluation rules}

\begin{itemize}
    \item Participants are free to develop their own attacker systems, using components of the provided baseline or not. 
     They are encouraged (but not required) to submit results for each anonymization system and to design attacker systems that target the specific weaknesses of each anonymization system.
    \item  Participants can use the training resources and development datasets specified in Section~\ref{sec:data} in order to train their system and tune hyperparameters.

    \item 
To compute the score for the pair \{set of enrollment utterances, trial utterance\} only the utterances included in this pair can be used from the evaluation data.

    \item Anonymization system authors and members of their team are welcome to participate. Their results will be considered as official only when attacking other systems than their own.
    If the authors release the code for the corresponding anonymization system before the deadline for \textit{“Declaration of  additional training/development data and models”} (see Table~\ref{tab:dates}), they can also attack their own anonymization system and participate in official ranking for this system.

\end{itemize}

\section{Registration and submission of results}

\paragraph{Registration.}

Participants/teams are requested to register for the evaluation.  Registration should be performed \textbf{once only} for each participating entity using the  
\href{https://forms.office.com/r/MTcyM4vj75}{registration form}.
Participants will receive a confirmation email within 48 hours after successful registration, otherwise or in case of any questions they should contact the organizers:

 \href{mailto:attacker.challenge@inria.fr?subject=Attacker registration}{attacker.challenge@inria.fr}.

\paragraph{Submission of results.}\label{subsec:submission}

Each participant may submit scores/results for one or more attacker systems, each targeting all anonymization systems or only some of them. 
Each single submission 
should include
  the EER  and  corresponding ASV scores 
     (for the development and evaluation data) obtained with the proposed attacker system for 4 trial lists (in the same format as generated by the baseline attacker system\footnote{Example: \href{https://www.dropbox.com/scl/fo/zwekrszawd0k2r2ahcuw4/ABkUi9qm0PRvmedQJln3dcM?rlkey=s0a3vx761zckxp4lpuhcaazhh&st=v0j14btm&dl=0}{link}}):
    \begin{itemize}
    {
    \setlength{\parskip}{0pt} \setlength{\itemsep}{0pt plus 1pt}
     
        \item  \textcolor{darkspringgreen}{data/libri\_dev\_trials\_f/trials} (example: \textcolor{darkspringgreen}{libri\_dev\_enrolls-libri\_dev\_trials\_f} \href{https://www.voiceprivacychallenge.org/attacker/data/libri_dev_enrolls-libri_dev_trials_f/scores}{scores})  
        
       \item  \textcolor{darkspringgreen}{data/libri\_dev\_trials\_m/trials} (example: \textcolor{darkspringgreen}{libri\_dev\_enrolls-libri\_dev\_trials\_m} \href{https://www.voiceprivacychallenge.org/attacker/data/libri_dev_enrolls-libri_dev_trials_m/scores}{scores})  

         \item \textcolor{darkspringgreen}{data/libri\_test\_trials\_f/trials} (example: \textcolor{darkspringgreen}{libri\_test\_enrolls-libri\_test\_trials\_f} \href{https://www.voiceprivacychallenge.org/attacker/data/libri_test_enrolls-libri_test_trials_f/scores}{scores})
       
        \item \textcolor{darkspringgreen}{data/libri\_test\_trials\_m/trials} (example: \textcolor{darkspringgreen}{libri\_test\_enrolls-libri\_test\_trials\_m} \href{https://www.voiceprivacychallenge.org/attacker/data/libri_test_enrolls-libri_test_trials_m/scores}{scores}) 
         }
 \end{itemize}

    All data should be submitted in the form of a single compressed archive.

Each participant should also submit a single, detailed system description.  
All submissions should be made according to the schedule below. Submissions received after the deadline will be marked as `late' submissions, without exception.
System descriptions will be made publicly available on the Challenge website.
Further details concerning the submission procedure 
will be published via the participants mailing list and via the \href{https://www.voiceprivacychallenge.org/attacker/}{VoicePrivacy Attacker Challenge website}.

\paragraph{Special session at ICASSP 2025.}

Results will be presented at the \href{https://2025.ieeeicassp.org/}{ICASSP 2025} special session to which 5 selected top-ranked participants will be invited to submit and present their challenge systems. All 
participants will be invited to submit the extended versions of their papers to the  SPSC 2025 Symposium.
Accepted papers will be published in the ICASSP proceedings. 
%
%
According to \url{https://2025.ieeeicassp.org/call-for-gc-proposals/}:
\textit{“The review process is coordinated by the challenge organizers and the SPGC chairs. 
All 2-page papers should be covered by an ICASSP registration and should be presented in person at the conference.”
}

\section{Schedule}\label{sec:schedule}

The result submission deadline is \colorbox{navajowhite}{\textbf{5th December 2024}}.

\begin{table}[tbh]
  \caption{Important dates}\label{tab:dates}
  \centering
   \resizebox{\textwidth}{!}{
   \resizebox{0.65\textwidth}{!}{
  \begin{tabular}{l r }
    \toprule
 Release of the  training, development and evaluation data, baselines and evaluation software & \textcolor{blue}{September 2024} \\ \midrule

  Declaration of  additional training/development data and models & \textcolor{blue}{13th October 2024} \\  \midrule
 Publication of the full final list of training data and models  & \textcolor{blue}{15th October 2024} \\ \midrule
 Deadline for participants to submit  scores, evaluation results  and system descriptions & \textcolor{blue}{5th December 2024} \\  \midrule 
  Deadline for participants to submit  2-page papers to ICASSP-2015 (by invitation only) & \textcolor{blue}{9th December 2024} \\  \midrule 
  Paper Acceptance Notification & \textcolor{blue}{30th December 2024} \\  
 \bottomrule
   \end{tabular}}
   }
\end{table}

\section{Acknowledgement}
This work was supported by the French National Research Agency under project Speech Privacy and project IPoP of the Cybersecurity PEPR.

\bibliographystyle{IEEEtran}
\bibliography{main}

\end{document}